\documentclass[graybox]{svmult}

\usepackage{mathptmx}       % selects Times Roman as basic font
\usepackage{helvet}         % selects Helvetica as sans-serif font
\usepackage{courier}        % selects Courier as typewriter font
\usepackage{type1cm}        % activate if the above 3 fonts are
\usepackage[utf8]{inputenc}
\usepackage[T1]{fontenc}
\usepackage{microtype}
\usepackage{makeidx}         % allows index generation
\usepackage{graphicx}        % standard LaTeX graphics tool
\usepackage{multicol}        % used for the two-column index
\usepackage[bottom]{footmisc}% places footnotes at page bottom

\usepackage{amsmath}

\usepackage{xcolor}

\makeindex             % used for the subject index
\usepackage{listings}

\usepackage[bookmarks=false]{hyperref}

\begin{document}

\title*{	
The SysML/KAOS Domain Modeling Approach }
\author{Steve TUENO, Régine LALEAU, Amel MAMMAR, Marc FRAPPIER}
\institute{Steve TUENO \at Université Paris-Est Créteil, 94010, CRÉTEIL, France, \at Université de Sherbrooke, Sherbrooke, QC J1K 2R1, Canada, \\\email{steve.tuenofotso@univ-paris-est.fr}
\and Régine LALEAU \at Université Paris-Est Créteil, 94010, CRÉTEIL, France, \email{laleau@u-pec.fr}
\and Amel MAMMAR \at Télécom SudParis, 91000, Evry, France, \email{amel.mammar@telecom-sudparis.eu}
\and Marc FRAPPIER \at Université de Sherbrooke, Sherbrooke, QC J1K 2R1, Canada, \email{Marc.Frappier@usherbrooke.ca}
}
\maketitle

\section{Introduction}\label{sec:1}
A means of building safe critical systems consists of formally modeling the requirements formulated by stakeholders and ensuring their consistency with respect to application domain properties.
 This paper proposes a metamodel for an ontology modeling formalism based on \textit{OWL} and \textit{PLIB}. This modeling formalism is part of a method for modeling the domain of systems whose requirements are captured through \textit{SysML/KAOS}.  The formal semantics of \textit{SysML/KAOS} goals are represented using \textit{Event-B} specifications.  Goals provide the set of events, while  domain models will provide the structure of the system state of the \textit{Event-B} specification. Our proposal is illustrated through a case study dealing with a \textit{Cycab} localization component specification \cite{anr_tacos_reference_link}. The case study deals with the specification of a localization software component that uses GPS, Wi-Fi and sensor technologies for the realtime localization of the \textit{Cycab} vehicle \cite{DBLP:conf/iros/SekhavatH00}, an autonomous ground transportation system designed to be robust and completely independent. 
%The mapping between our domain model representation and \textit{Event-B} is out of the scope of this paper.

The remainder of this paper is structured as follows: Section 2 briefly describes the \textit{SysML/KAOS} method.
 Follows a presentation, in Section 3,  of the relevant state of the art on domain modeling in 
%the wake of 
requirements engineering and a comparison of ontology modeling formalisms. In Section 4, we describe and illustrate our approach to model the domain of a system specified using the \textit{SysML/KAOS} method.

 \section{SysML/KAOS}
Requirements engineering focuses on defining and handling requirements. These and all related activities, in order to be carried out, require the choice of an adequate means for  requirements representation. The \textit{KAOS} method \cite{DBLP:books/daglib/0025377,DBLP:conf/isola/MammarL16},
% whose reasons for choosing are explained in \cite{DBLP:conf/isola/MammarL16},
   proposes to represent the requirements in the form of goals, which can be \textit{functional} or \textit{non-functional}, through five sub-models of which the two main ones are : 
 \textbf{the object model} which uses the \textit{UML} class diagram for the representation of domain  vocabulary and \textbf{the goal model} for the determination of  requirements to be satisfied by the system and of expectations with regard to the environment through a goals hierarchy having strategic goals formulated by  stakeholders at the root level.
% allowing to derive the latter from  strategic objectives formulated by the various stakeholders.
The hierarchy is built through a succession of refinements using different operators : \textbf{\textit{AND}}, \textbf{\textit{OR}} and \textbf{\textit{MILESTONE}}.
An \textit{\textbf{AND refinement}} decomposes a goal into subgoals, and all of them must be achieved to realise the parent goal.  Dually, an \textit{\textbf{OR refinement}} decomposes a goal into subgoals
such that the achievement of only one of them is sufficient for the accomplishment of the parent goal. A \textbf{\textit{MILESTONE refinement}} is a variant of \textit{\textbf{AND refinement}} which allows the definition of an achievement order  between goals.
 Requirements and expectations correspond to the lowest level goals of the  model.
 %\textbf{The responsibility, the operationnalization    and  the behavioural  models} are respectively used for the representation of assignment links between requirements and system agents and between expectations and environmental agents,  for the representation of  operations associated with the implementation of requirements  and for the materialization of  links between agents and operations.

%A \textit{KAOS}  goal may be \cite{DBLP:conf/iceccs/MatoussiGL11} :
%\begin{itemize}
%\item  \textbf{Achieve } with for dual \textbf{Cease} for guaranteeing reaching a target state for the system from a certain current state. 
%\item \textbf{Maintain } with for dual \textbf{Avoid} for the guarantee of the maintenance of a certain state once it is reached.  
%\end{itemize}

\textit{KAOS} proposes a structured approach to obtaining the requirements based on  expectations formulated by  stakeholders. Unfortunately, it  offers no mechanism to maintain a strong traceability between   those requirements and  deliverables associated with system  design and implementation, making it difficult to validate them against the needs formulated.
The \textit{SysML UML profile} has been  specially designed by the Object Management Group (OMG)  for the analysis and specification of complex systems and allows for the capturing of requirements and the maintaining of   traceability links between those requirements and  design diagrams resulting from the system design phase.
Unfortunately,  OMG has not defined a formal semantics and an unambiguous syntax for requirements specification. \textit{SysML/KAOS} \cite{DBLP:conf/inforsid/GnahoS10} therefore proposes to extend the \textit{SysML} metamodel with a set of concepts allowing to represent  
% functional and non-functional
  requirements in \textit{SysML} models as \textit{KAOS}  goals.
 %, which makes their representation more precise and more formal

%%%%%%%%% SYSML/KAOS %%%%%%%%%%%%%%%%
A functional goal, under \textit{SysML/KAOS}, describes the \textit{expected behaviour} of the system once a certain condition holds  \cite{DBLP:conf/isola/MammarL16} : \textit{[if \textbf{CurrentCondition} then] sooner-or-later \textbf{TargetCondition}}.
%\textit{\textbf{Example: } if handle is down then landing gear must be extended}.  \\
\textit{SysML/KAOS} allows the definition of a functional goal without specifying a \textit{CurrentCondition}. In this case,  the expected behaviour can be observed  from any system state. 
%\textit{\textbf{Example: } put handle down. } 

%In the same throw, non-functional goals are defined by specifying, for each of them, the targeted property and the concerned system entity \cite{DBLP:journals/isi/GnahoS11}. \textit{\textbf{Example: } \textbf{\textit{Availability [Vehicle Localization]}} is concerned with the availability of the vehicle localization}.  
%In the remainder of this paper, we will only focus on functional requirements. \\

\begin{figure}[!h]
\begin{center}
\includegraphics[width=1.1\textwidth]{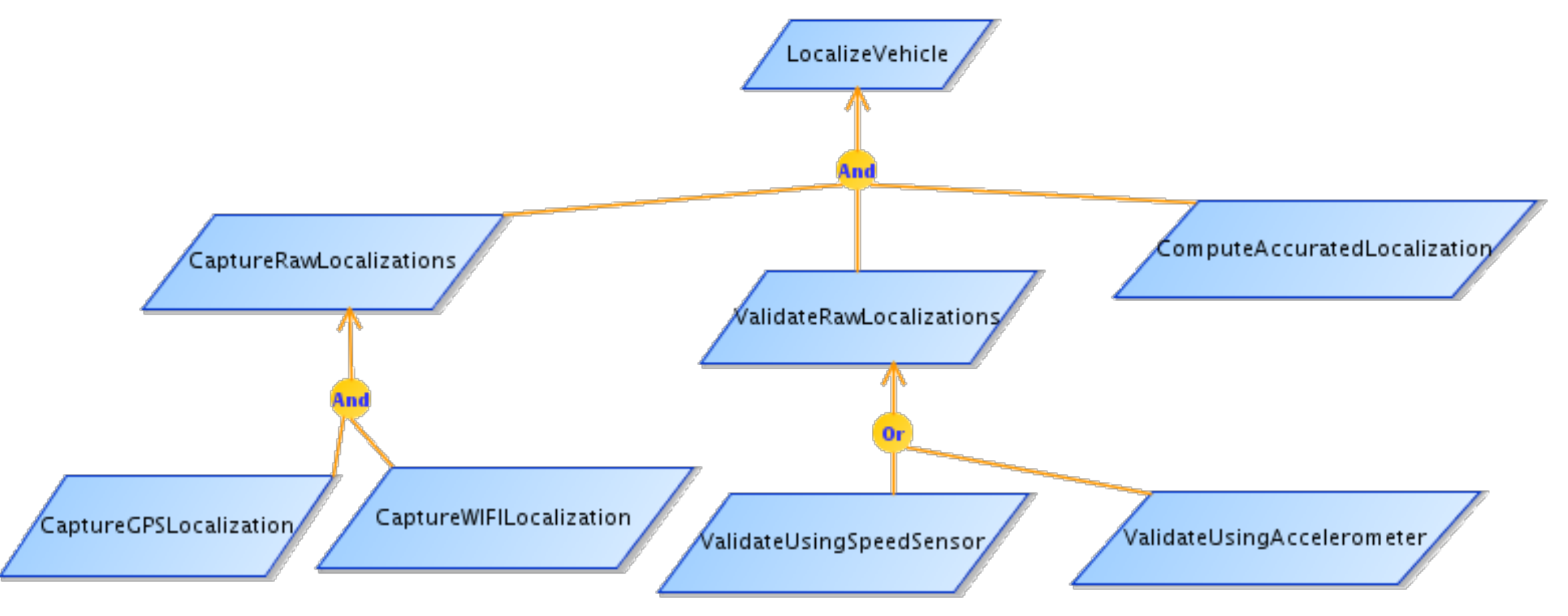}
\end{center}
\caption{\label{localization_component_goal_diagram} Excerpt from the localization component goal diagram}
\end{figure}

Figure \ref{localization_component_goal_diagram} is a goal diagram from the \textit{Cycab System} localization component   focused on the purpose of  vehicle localization.

To achieve the root goal, which is the localization of the vehicle (\textbf{LocalizeVehicle}), raw localizations must be captured from vehicle sub components (\textbf{CaptureRawLocaliza\-tions}) which can be GPS  (\textbf{CaptureGPSLocalization}) or Wi-Fi (\textbf{CaptureWIFILocaliza\-tion}),  be validated using a vehicle sensor  (\textbf{ValidateRawLocalizations}) which has to be  either a speed sensor  (\textbf{ValidateUsingSpeedSensor}) or an accelerometer (\textbf{ValidateUsingAccelerometer})  and used to compute the vehicle's accurate localization (\textbf{ComputeAccuratedLocalization}).

\section{State of the Art On Domain Modeling in Requirements Engineering}
\label{sec:3}

\subsection{Existing Approaches}

In \textit{KAOS} \cite{DBLP:books/daglib/0025377}, 
%the representation of the domain of a system called \textit{object model} is made using \textit{UML} class diagrams. 
the domain of a system is specified by an \textit{object model} described by \textit{UML} class diagrams. An object within this model can be an \textbf{\textit{entity}} if it exists independently of the others and does not influence the state of any other object, an \textbf{\textit{association}} if it links other objects on which it depends, an \textbf{\textit{agent}} if it actively influences the system state by acting on other objects or an \textbf{\textit{event}} if its existence is instantaneous, appearing to impulse an update of the system state. 
This approach, which is essentially graphic and informal as argued in \cite{DBLP:conf/icse/McUmberC01}, is difficult to exploit in
case of critical systems \cite{DBLP:journals/sqj/NguyenVLG14}. Moreover, it does not offer mechanisms for referencing a model within another, which limits the reusability of models.

% In \cite{zong1domain}, domain is modeled using \textit{description logic}.

In \cite{DBLP:journals/icae/Devedzic01}, author proposes to model the knowledge of the domain through either formulae of first-order logic or ontologies.
He considers an ontology as a more structured and extensible representation of domain knowledge.
%He ends up concluding that the use of predicate logic cannot be generalized.

In \cite{DBLP:conf/icsoft/KitamuraHKS07},
the domain model is built around the notions of \textit{Concept} and \textit{Relationship}.
Each entry in this model consists of an assertion linking two instances of \textit{Concept} through a \textit{Relationship} instance.
A categorization is proposed for concepts and relationships : a concept can be a function, an object, a constraint, an actor, a platform, a quality or an ambiguity, while a relationship can be a performative  or a symmetry, reflexivity or transitivity relation. However, the proposed metamodel appears to be incomplete. Indeed, it does not allow to represent key elements of  domain modeling  as for example the cardinality of a relationship or the attributes of a concept. Moreover, it does not allow to establish references between several models, which limits their reuse.

In \cite{DBLP:journals/sqj/NguyenVLG14}, ontologies are used not only to represent domain knowledge, but also to model and analyze requirements. The proposed methodology is called \textit{knowledge-based requirements engineering (KBRE)} and is mainly used for  detection and processing of inconsistencies, conflicts and redundancies among requirements. In spite of the fact that \textit{KBRE} proposes to model the domain knowledge  through ontologies, the proposal focuses on the representation of requirements and proposes nothing regarding  domain modeling.   It is in the same vein that the \textit{GOORE method}  is presented in \cite{DBLP:conf/er/ShibaokaKS07}.

In \cite{DBLP:journals/re/DermevalVBCIBS16}, authors are interested in a systematic review of the literature related to applications of ontologies in requirements engineering. They end up describing ontologies  as a standard form of formal representation of concepts within a domain, as well as of relationships between those concepts. This is equivalent to considering ontologies as a standard for formal modeling of system domain.\\

These approaches suggest that ontologies are relevant for modeling the domain of a system.

\subsection{A Study of Ontology Modeling Formalisms}
An ontology can be defined as a formal model representing concepts that can be grouped into categories through generalization/specialization relations, their instances, constraints and properties as well as relations existing between them. Ontology modeling formalisms can be grouped into two categories : \textit{Closed World Assumption (CWA)} for those considering that any fact that cannot be deduced from what is declared is false and \textit{Open World Assumption (OWA)} for those considering
that there may be facts that cannot be deduced from what is specified and that can be true. As \cite{DBLPjournals/soco/AmeurBBJS17}, we consider that accurate modeling of the knowledge of engineering domains, to which we are interested, must be done under the \textit{CWA} assumption. Indeed, this assumption improves the formal validation of the consistency of system’s specifications with respect to domain properties. Moreover, systems of interest to us are so critical that no assertion should be assumed to be true until consensus is reached on its veracity. Similarly, we also advocate \textit{strong typing} \cite{DBLPjournals/soco/AmeurBBJS17}, because our domain models are made in order to complete \textit{Event-B} models for the system specifications to be formally validated.

Several ontology formalisms exist. The main ones are \textbf{\textit{OWL (Ontology Web Language)}} \cite{DBLP:reference/snam/SenguptaH14}, \textbf{\textit{PLIB (Part LIBrary)}} \cite{DBLP:conf/ifip/Pierra04} and \textbf{\textit{F-Logic (Frame Logic)}} \cite{DBLP:conf/sigmod/KiferL89}. A summary of the similarities and differences between these  ontology modeling formalisms is presented through Table \ref{tableau_recapitulatif_conclusions_comparaison_owl-vs-plib-F-Logic}:

\begin{table}[htb]
\begin{center}
\caption{\label{tableau_recapitulatif_conclusions_comparaison_owl-vs-plib-F-Logic} Comparative table of the three main ontology modeling formalisms}
\begin{tabular}{|p{5cm}|p{2cm}|p{2.5cm}|p{2cm}|}
\hline
\textbf{Characteristics} & \textbf{ OWL} & \textbf{PLIB} & \textbf{F-Logic} \\
\hline
\textbf{Modularity} & total & \textbf{ partial} & total \\
\hline
\textbf{CWA vs OWA } & OWA & \textbf{CWA} &  \textbf{CWA} \\
\hline
\textbf{Inheritance} & multiple & \textbf{simple} & multiple \\
\hline
\textbf{Typing} & weak & \textbf{strong {\tiny (any element belongs to one and only one type)}} & weak \\
\hline
\textbf{Contextualization of a property {\footnotesize (parameterized attributes)}} & - & \textbf{+} & \textbf{+} \\
\hline
\textbf{Different views for an element} & - & \textbf{+} & - \\
\hline
\textbf{Graphic representation} & \textbf{+} & - & - \\
\hline
\textbf{Domain Knowledge (static vs dynamic)} & static & static & static \\
\hline

\end{tabular}
\end{center}
\end{table}

\begin{itemize}
\item \textit{PLIB}, \textit{OWL} and\textit{ F-Logic} implement referencing mechanisms between ontologies. \textit{PLIB} supports partial import: a class of an ontology \textit{A} can extend a class of an ontology \textit{B} and explicitly specify the properties it wishes to inherit. Moreover, if nothing is specified, no property will be imported. On the other hand, \textit{OWL} and \textit{F-Logic} use the total import: when an ontology \textit{A} refers to an ontology \textit{B}, all the elements of \textit{B} are  accessible within \textit{A}.
\item  \textit{PLIB} and \textit{F-Logic} use the \textit{CWA} assumption for constraint verification, \textit{OWL} uses the \textit{OWA} assumption. %Par contre, plusieurs propositions de sémantiques et de raisonneurs OWL avec l'approche CWA existent.
\item \textit{OWL} and \textit{F-Logic } implement multiple inheritance and instantiation. \textit{PLIB} implements simple inheritance and  instantiation.
On the other hand, with the \textit{\textbf{is\_case\_of}} relation, a \textit{PLIB} class can be \textit{a case of} several other classes, each class bringing some specific properties.
\item \textit{PLIB} is strongly typed (any element belongs to one and only one type), which is not the case for \textit{OWL} and \textit{F-Logic}.
\item  \textit{PLIB} and \textit{F-Logic} allow the definition of parameterized attributes using context parameters, which is not possible with \textit{OWL}.
\item \textit{PLIB} allows the association of several representations or view points with a concept, which is not possible with neither \textit{OWL} nor\textit{ F-Logic}.
\item \textit{OWL}, \textit{PLIB} and \textit{F-Logic} are focused only on modeling of static domain knowledge. It is for example impossible to specify that the localization of a vehicle can change dynamically while its brand can't.

\end{itemize}

We can observe, as stated in \cite{zong1domain}, that   \textit{"unfortunately, all the studied formalisms emphasize more on modeling static domain knowledge"}. None of these formalisms allows  to specify that a knowledge described must remain unchanged or that it is likely to be updated. 
Moreover, the construction of an \textit{OWL} ontology is done under the \textit{OWA} assumption and \textit{PLIB} does not allow the specification of rules allowing to deduce new facts from existing ones. Finally, \textit{F-Logic} as \textit{OWL} are weakly typed.

\section{Our Approach for Domain Modeling}\label{sec:4}
We have chosen to represent domain knowledge using ontologies since they are semantically richer and therefore allow a more explicit representation of domain characteristics. Thus, in this part, we propose a metamodel, based on that of \textit{OWL} and \textit{PLIB} and  conforming to the \textit{CWA assumption}  for the representation of the domain of a system whose requirements are captured using the  \textit{SysML/KAOS} method.
Our formalism make the \textit{Unique Name Assumption (UNA)} \cite{DBLPjournals/soco/AmeurBBJS17} : the name of an element is sufficient to uniquely identify it among all the others within a domain model.
%;  if two elements have the same name, then they are identical. 
 Furthermore, our metamodel is designed to allow the specification  of   knowledge that is likely to evolve over time. 
%MODIFIER POUR PRECISER QUE LE METAMODELE EN EST ENCORE A SES PREMICES ET QU'IL EST SUCCEPTIBLE D'ÊTRE ETENDU/ENRICHI (simplement).
 %It may be extended to cover specific use cases. 
 We have identified two graphical syntaxes for the representation of ontologies : the syntax proposed by \textit{OntoGraph} \cite{protege_ontograph_reference_link} and the syntax proposed by   \textit{OWLGred} \cite{owlgred_reference_link}.
The \textit{OntoGraph} syntax is the one used in \cite{DBLP:conf/isola/MammarL16}. Unfortunately, it does not allow the representation of some domain model elements such as attributes or cardinalities. For our case study, we have thus decided to use the \textit{OWLGred} syntax. 

%Furthermore, for readability purposes, we have decided to remove optional characteristics representation and to represent the \textit{isVariable} attribute only when it is set to \textit{true}: when \textit{isVariable} is set to \textit{true} for an element, it is tagged with the stereotype \textbf{\textit{<<isVariable>>}}.

%Sub-section \ref{our-approach-presentation} deals with the abstract syntax associated with our domain modeling formalism and  Sub-section \ref{our-approach-illustration} with its current concrete syntax.

%\subsection{Presentation}\label{our-approach-presentation}

We present through Figures \ref{our_businessdomain_metamodel_part1}, \ref{our_businessdomain_metamodel_part4}, \ref{our_businessdomain_metamodel_part2} and \ref{our_businessdomain_metamodel_part3}  
%and \ref{our_businessdomain_metamodel_bis}
 the main part of the metamodel associated with our domain modeling approach, knowing that yellow elements are those having an equivalent in \textit{OWL} metamodel and that red ones are those that we have either  inserted or customized. Furthermore, some constraints and associations, such as the \textit{parentConcept} association, have been extracted from the \textit{PLIB} metamodel.
Due to space consideration, we will not highlight all the elements and constraints of the metamodel. Figures \ref{localization_component_shonan_book_ref0}, \ref{localization_component_shonan_book_ref2} and \ref{localization_component_shonan_book_ref1}  represent respectively the domain model associated to the root level of the \textit{SysML/KAOS} goal diagram illustrated through Figure   \ref{localization_component_goal_diagram}, that associated with the second level of refinement and that associated with the first one. 
The domain model associated to the goal diagram root level is named \textit{"untitled-ontology-52"}, the one associated to the first refinement level is named \textit{"untitled-ontology-53"} and the one associated to the second refinement level is named \textit{"untitled-ontology-54"}.

\subsection{Concepts and Individuals,  Data Sets and Data Values}

\begin{figure}[!h]
\begin{center}
\includegraphics[width=1\textwidth]{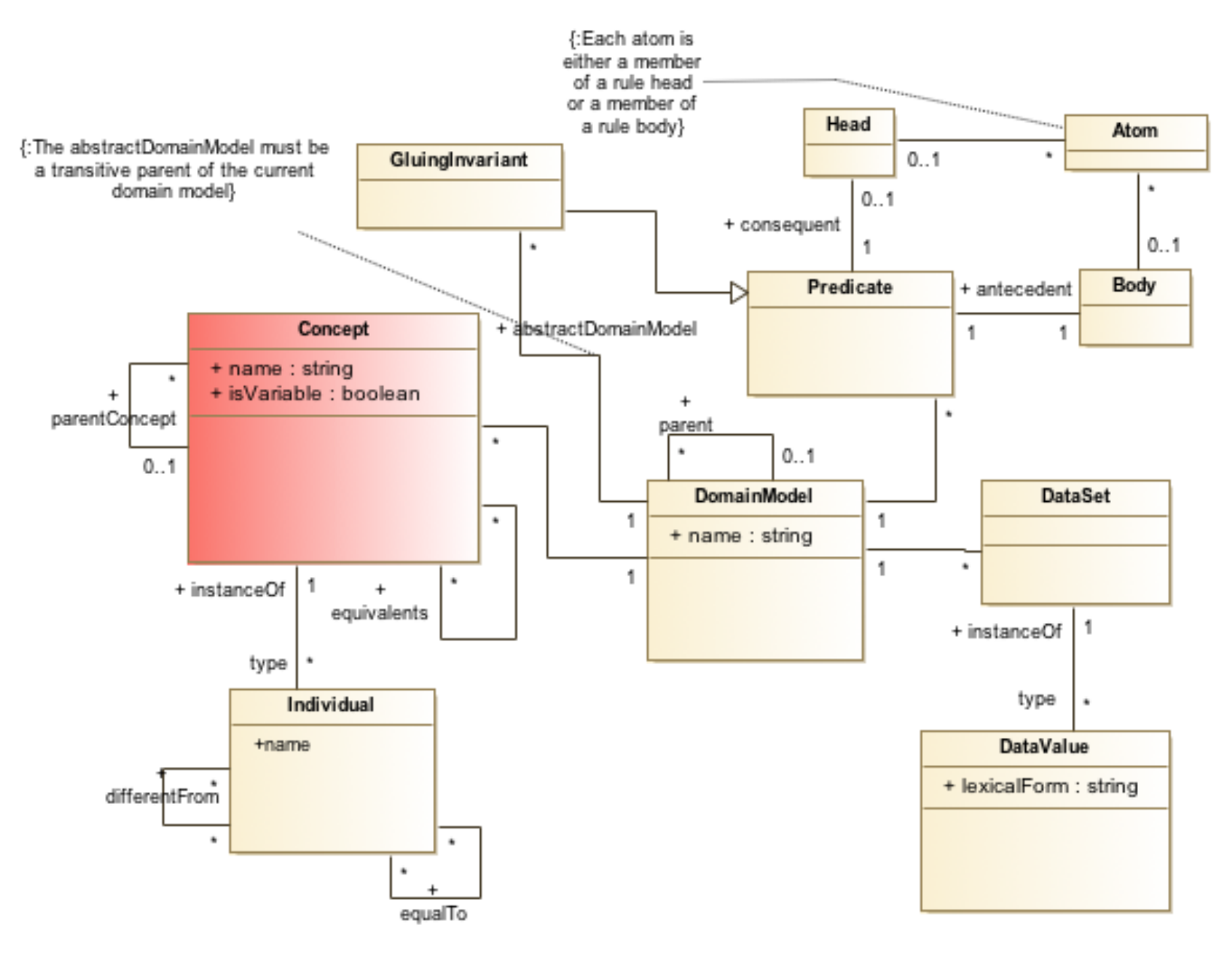}
\end{center}
\caption{\label{our_businessdomain_metamodel_part1} First part of the metamodel associated with domain modeling}
\end{figure}

\begin{figure}[!h]
\begin{center}
\includegraphics[width=1\textwidth]{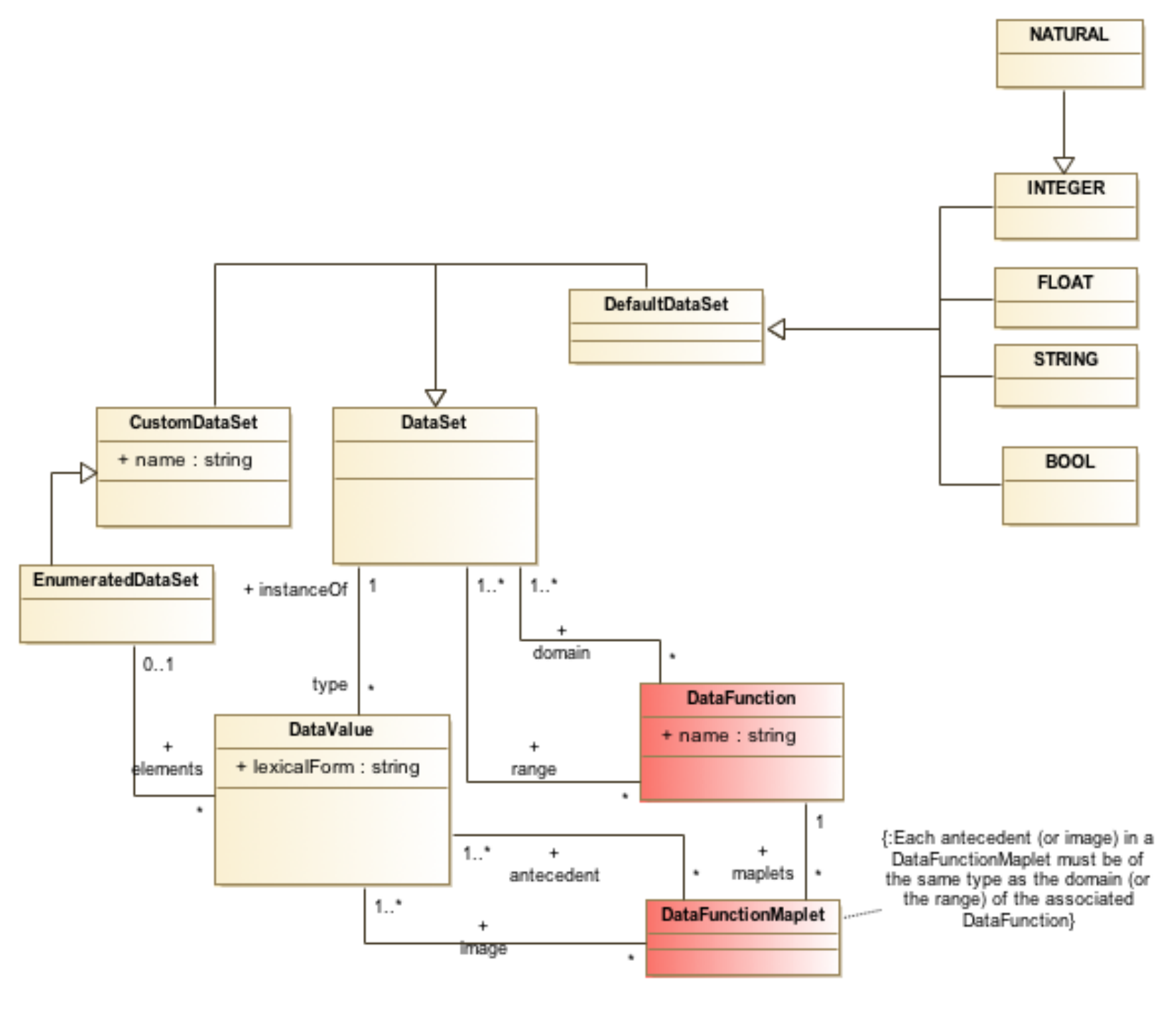}
\end{center}
\caption{\label{our_businessdomain_metamodel_part4} Fourth part of the metamodel associated with domain modeling}
\end{figure}

\begin{figure}[!h]
\begin{center}
\includegraphics[width=1.2\textwidth]{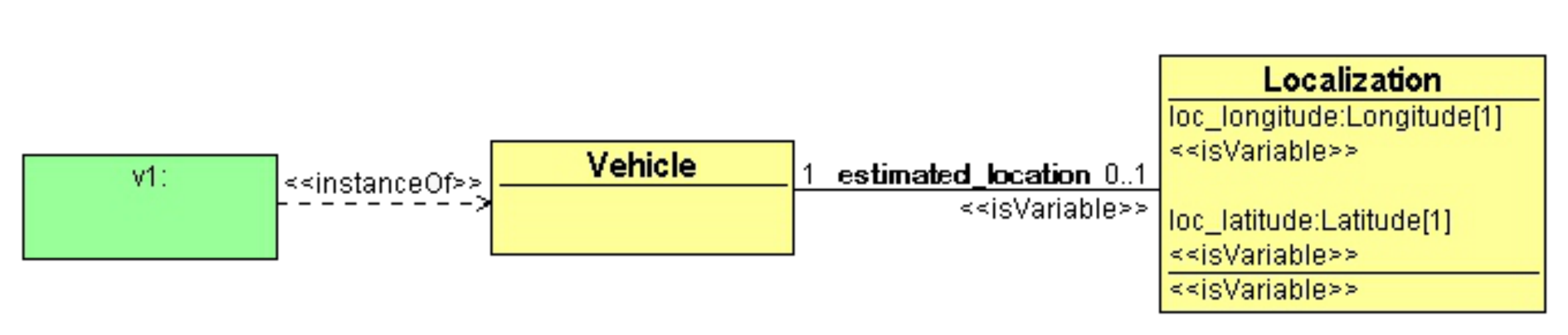}
\end{center}
\caption{\label{localization_component_shonan_book_ref0} \textit{\textbf{untitled-ontology-52}}: ontology associated to the root level}
\end{figure}

 The central notion is the notion of \textit{Concept} which  represents a group of individuals sharing common characteristics (Fig. \ref{our_businessdomain_metamodel_part1}). 
A \textit{concept} can be declared \textit{variable} (\textit{isVariable=true}) when the set of its individuals is likely to be updated through  addition or deletion of individuals. Otherwise, it is considered to be \textit{constant} (\textit{isVariable=false}).
A concept may be associated with another, known as its parent concept, through the \textit{parentConcept} association, from which it inherits   properties. As a result, any individual of the child concept is also an individual of the parent concept. 

In \textit{untitled-ontology-52} (Fig. \ref{localization_component_shonan_book_ref0}), a Vehicle is modeled as an instance of \textit{Concept} named \textit{"Vehicle"} and its localization is represented through an instance of \textit{Concept} named \textit{"Localization"}. For readability purposes, we have decided to represent the \textit{isVariable} attribute only when it is set to \textit{true}. Since it is possible to dynamically add or remove localizations, the attribute \textit{isVariable} of \textbf{\textit{Localization}} is  set to \textit{true}, which is represented by the stereotype \textit{<<isVariable>>}. Since the system is designed to control a single vehicle, it is not possible to dynamically add new ones. The involved vehicle is  thus modeled as an instance of \textit{Individual} named \textit{"v1"} having \textbf{\textit{Vehicle}} as \textit{type}.

An instance of \textit{DataSet} is used to group instances of \textit{DataValue} having  the same type (Fig. \ref{our_businessdomain_metamodel_part4}). Default \textit{DataSets} are      \textit{INTEGER},  \textit{NATURAL} for positive integers,  \textit{FLOAT},  \textit{STRING} or  \textit{BOOL} for booleans. The most basic way to build an instance of \textit{DataSet} is by listing its elements. This can be done through the \textit{DataSet} specialization   called \textit{EnumeratedDataSet}.

\subsection{Relations and   Attributes}

\begin{figure}[!h]
\begin{center}
\includegraphics[width=1\textwidth]{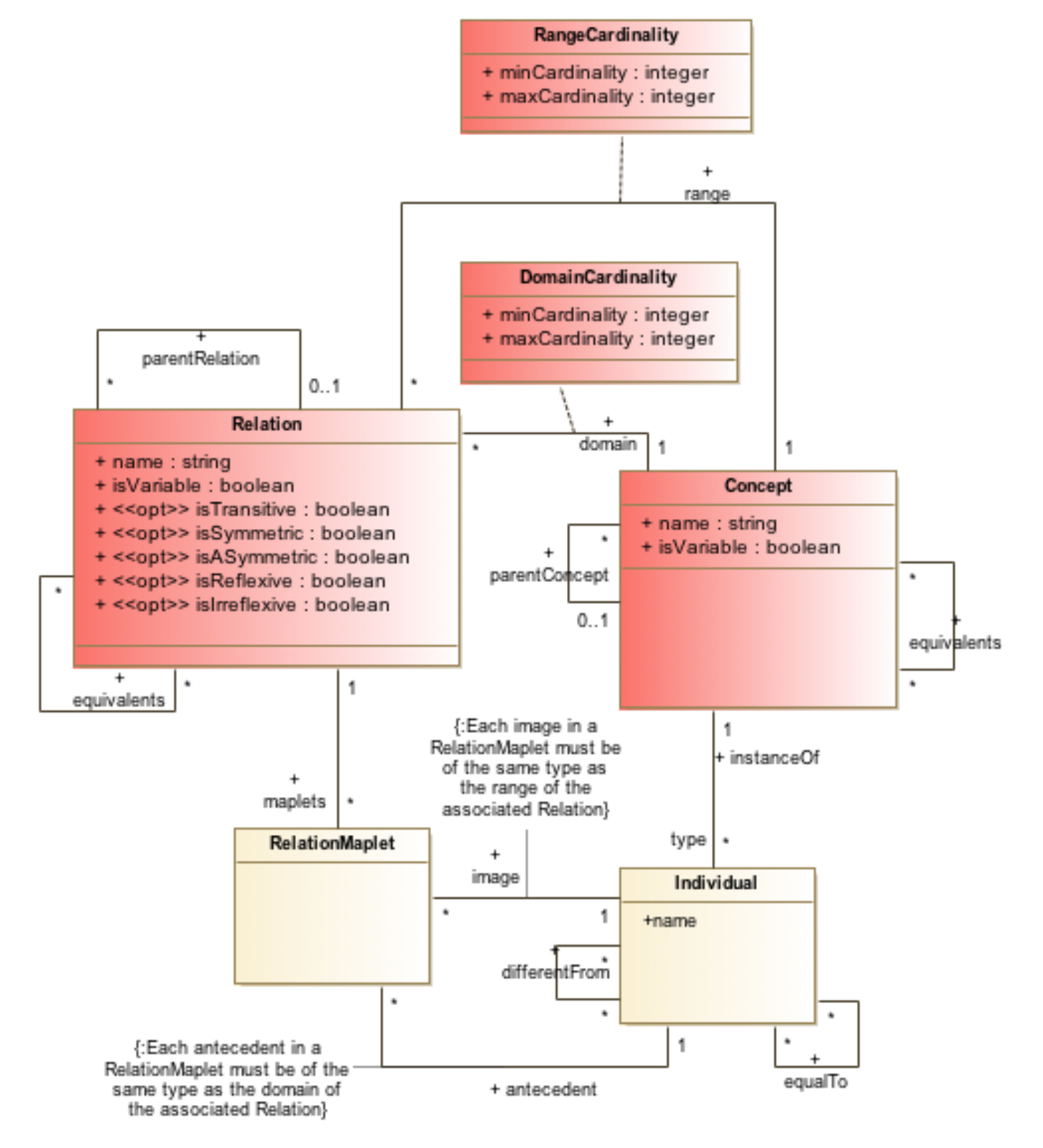}
\end{center}
\caption{\label{our_businessdomain_metamodel_part2} Second part of the metamodel associated with domain modeling}
\end{figure}

\begin{figure}[!h]
\begin{center}
\includegraphics[width=1\textwidth]{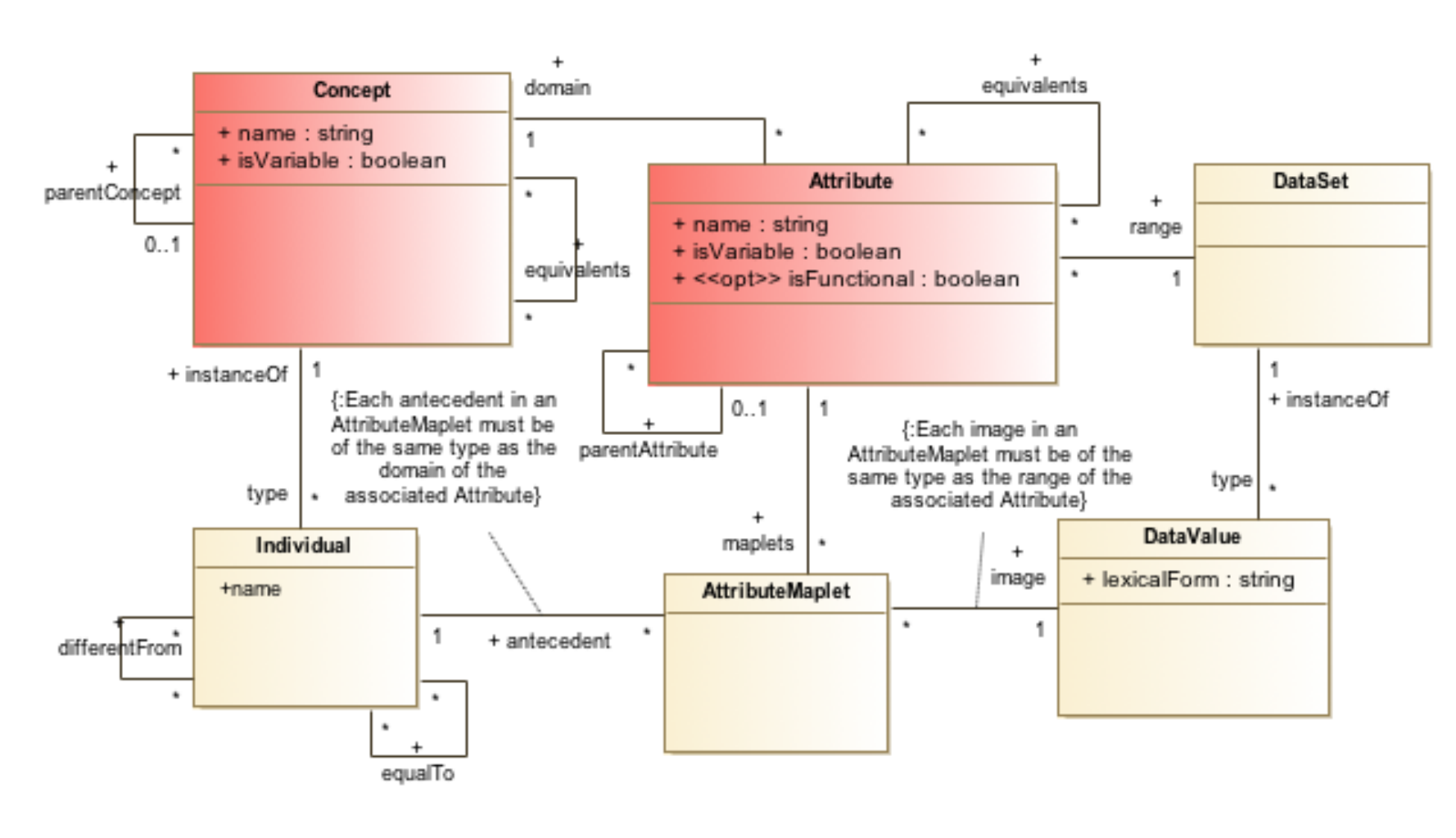}
\end{center}
\caption{\label{our_businessdomain_metamodel_part3} Third part of the metamodel associated with domain modeling}
\end{figure}

The notion of \textit{Relation} is used to capture  links between concepts (Fig. \ref{our_businessdomain_metamodel_part2}) and the notion of \textit{Attribute}  links between concepts and data sets (Fig. \ref{our_businessdomain_metamodel_part3}).
A relation (Fig. \ref{our_businessdomain_metamodel_part2})  or an attribute (Fig. \ref{our_businessdomain_metamodel_part3}) can be declared \textit{variable} if the list of maplets related to it  is likely to change over time. Otherwise, it is considered to be \textit{constant}. 
The association between a relation and a concept is characterized by the \textit{cardinality} : \textit{DomainCardinality} and \textit{RangeCardinality} (Fig. \ref{our_businessdomain_metamodel_part2}).
Each instance of \textit{DomainCardinality} (respectively \textit{RangeCardinality}) makes it possible to define, for an instance of \textit{Relation} \textit{re}, the minimum and maximum limits of the number of instances of \textit{Individual}, having the domain (respectively range) of \textit{re} as \textit{type}, that can be put in relation with one instance of \textit{Individual}, having the range (respectively domain) of \textit{re} as \textit{type}.
 The following constraint is associated with these limits : $(minCardinality \geq 0) \wedge (maxCardinality = null \vee maxCardinality \geq minCardinality)$, knowing that if $maxCardinality=null$, then the maximum limit is \textit{infinity}.
Instances of \textit{RelationMaplet} are used to define associations between instances of \textit{Individual} through  instances of \textit{Relation}. In an identical manner, instances of \textit{AttributeMaplet} are used to define associations between instances of \textit{Individual}  and instances of \textit{DataValue} through  instances of \textit{Attribute}.

Optional characteristics can be specified for a relation  (Fig. \ref{our_businessdomain_metamodel_part2}) : \textit{transitive} (\textit{isTransitive}, default \textit{false}), \textit{symmetrical} (\textit{isSymmetric}, default \textit{false}), \textit{asymmetrical} (\textit{isASymmetric}, default \textit{false}), \textit{reflexive} (\textit{isReflexive}, default \textit{false}) or \textit{irreflexive} (\textit{isIrreflexive}, default \textit{false}). 
It is said to be \textit{transitive} (\textit{isTransitive=true}) when the relation of an individual \textit{x} with an individual \textit{y} which is in turn in relation to \textit{z} results in the relation of \textit{x} and \textit{z}.
It is said to be  \textit{symmetric} when the relation between an individual \textit{x} and an individual \textit{y} results in the relation of \textit{y} to \textit{x}.
It is said to be \textit{asymmetric} when the  relation of an individual \textit{x} with an individual  \textit{y} has the consequence of preventing a possible relation between \textit{y} and \textit{x}, with the assumption that $x \neq y$.
It is said to be \textit{reflexive} when every individual of the domain is in relation with itself.
It is finally said to be \textit{irreflexive} when it does not authorize any connection of an individual of the domain with itself.
Moreover, an attribute can be \textit{functional} (\textit{isFunctional}, default \textit{true}) if it associates to each individual of the domain one and only one data value of the range. 

For readability purposes, we have decided to remove optional characteristics representation and to represent the \textit{isVariable} attribute only when it is set to \textit{true}. 
In \textit{untitled-ontology-52} (Fig. \ref{localization_component_shonan_book_ref0}), \textbf{\textit{Localization}}  is the \textit{domain} of two attributes :  the latitude modeled as an instance of \textit{Attribute} named \textit{"loc\_latitude"} 
% represented through the concept \textbf{\textit{Latitude}}
 and the longitude modeled as an instance of \textit{Attribute} named \textit{"loc\_longitude"}.
\textbf{\textit{loc\_latitude}} has, as range, an instance of \textit{CustomDataSet} named \textit{"Latitude"} and \textbf{\textit{loc\_longitude}}  an instance of \textit{CustomDataSet} named \textit{"Longitude"}.
  Since it is possible to dynamically change the localization of a vehicle, the attribute \textit{isVariable} of  \textbf{\textit{loc\_latitude}} and that of \textbf{\textit{loc\_longitude}} are set to \textit{true}, which is represented by the stereotype \textit{<<isVariable>>}. The assocation between an instance of \textit{Vehicle} and an instance of \textit{Localization} is represented through an instance of \textit{Relation} named \textit{"estimated\_location"}. Its associated instance of \textit{DomainCardinality} has \textit{1} as \textit{minCardinality} and \textit{maxCardinality}, and its associated instance of \textit{RangeCardinality} has \textit{0} as \textit{minCardinality} and \textit{1} as  \textit{maxCardinality}.

\begin{figure}[!h]
\begin{center}
\includegraphics[height=0.84\textheight]{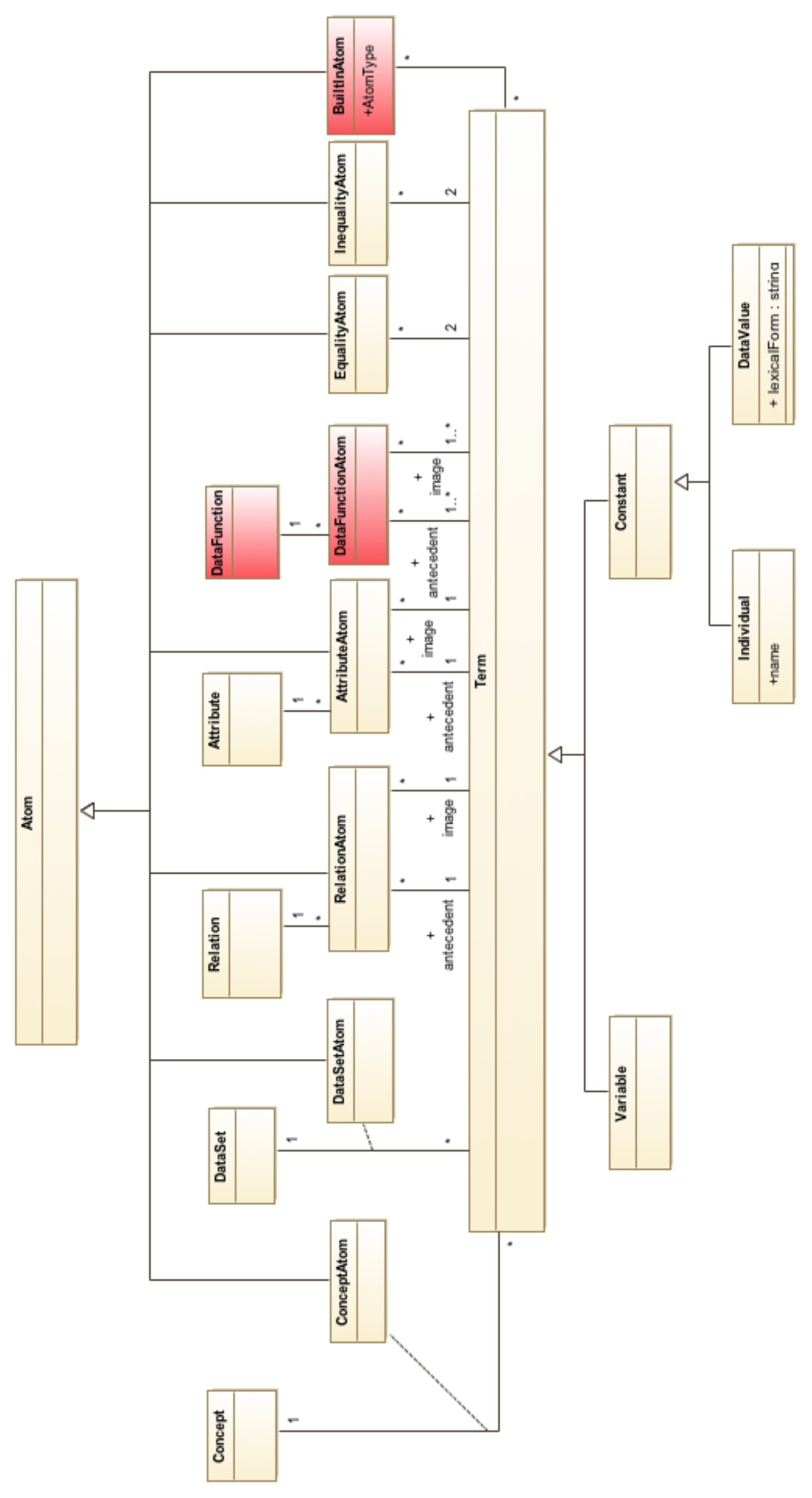}
\end{center}
\caption{\label{our_businessdomain_metamodel_bis} Fifth part of the metamodel associated with domain modeling}
\end{figure}

\subsection{Functions and Predicates}

 \begin{figure}[!h]
\begin{center}
\includegraphics[width=1.2\textwidth]{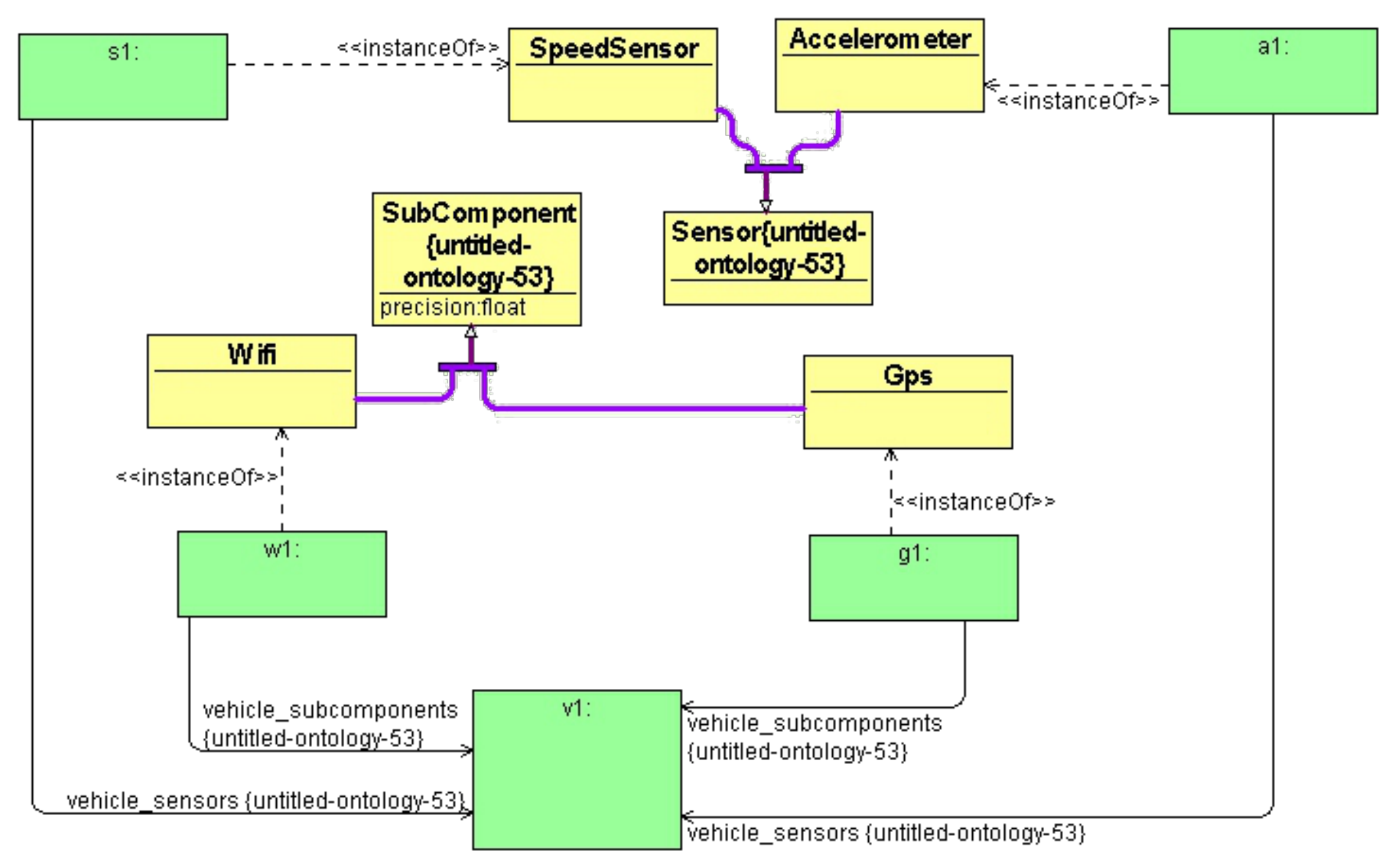}
\end{center}
\caption{\label{localization_component_shonan_book_ref2} \textit{\textbf{untitled-ontology-54}}: ontology associated to the second level of refinement}
\end{figure}

The notion of \textit{DataFunction} (Fig. \ref{our_businessdomain_metamodel_part4}) makes it possible to define operations which allow to determine data values at the output of a set of processes on some input data values. At each tuple of data values of the  domain, the \textit{data function} assigns a tuple of data values of the  range, and this assignement cannot be changed dynamically. \textbf{\textit{Example}}: We can define an instance of \textit{DataFunction} named \textit{"multiply"} to produce, given two instances of \textit{INTEGER} \textit{x} and \textit{y}, the instance of \textit{INTEGER} representing $x * y$.
On the other side, the notion of \textit{Predicate} (Fig. \ref{our_businessdomain_metamodel_part1})  is used to represent constraints between different elements of the domain model in the form of \textit{Horn clauses}: each predicate has a body which represents its \textit{antecedent} and a head which represents its \textit{consequent}, body and head designating conjunctions of atoms.
A \textit{typing atom} is used to define the type of a term : \textit{ConceptAtom} for individuals and \textit{DataSetAtom} for data values (Fig. \ref{our_businessdomain_metamodel_bis}). An \textit{association atom} is used to define associations between terms : \textit{RelationshipAtom} for the connection of two terms through a \textit{relation},  \textit{AttributeAtom} for the connection of two terms through an \textit{attribute} and  \textit{DataFunctionAtom } for the connection of terms through a \textit{data function} (Fig. \ref{our_businessdomain_metamodel_bis}). For each case, the types of terms must correspond to the domains/ranges of the considered link. A \textit{comparison atom} is used to define  comparison relationships between terms : \textit{EqualityAtom} for equality and \textit{InequalityAtom} for difference (Fig. \ref{our_businessdomain_metamodel_bis}). Built in atoms are some specialized atoms, characterized by identifiers captured through the \textit{AtomType} enumeration, and used for the representation of particular constraints between several terms (Fig. \ref{our_businessdomain_metamodel_bis}).
 For example, an arithmetic constraint between several integer data values.
 
%ILLUSTRATION AVEC LA PREDICAT DU CAS D'USAGE 
In \textit{untitled-ontology-54} (Fig. \ref{localization_component_shonan_book_ref2}), the constraint \textit{"a GPS is more precise than a Wi-Fi"}   is translated into an instance of  \textit{Predicate} represented through formula \ref{rule-1} : If an instance of \textit{Term}, named \textit{"x"}, having \textbf{\textit{Wifi}} as its \textit{type}, has \textbf{\textit{px}} as its \textit{precision} and an instance of \textit{Term}, named \textit{"y"}, having \textbf{\textit{Gps}} as its \textit{type}, has \textbf{\textit{py}} as its \textit{precision}, then $py>px$.
\begin{equation}
\label{rule-1}
greaterThan(?py, ?px) \leftarrow Wifi(?x) \wedge precision(?x, ?px) \wedge Gps(?y) \wedge precision(?y, ?py)  
\end{equation}
 
\textit{Predicates} can be used to \textit{parameterize}  relations or attributes in order to define dependent associations. For example, knowing that the resistance of a material depends on the temperature of the medium, resistance and temperature attributes are dependent.
%: \textbf{\textit{when}} the medium has a specific temperature, \textbf{\textit{then}} the material resistance has a specific value.
\textsf{GluingInvariant} (Fig. \ref{our_businessdomain_metamodel_part1}), specialization of \textsf{Predicate}, is used to represent links between variables elements defined within a domain model and those appearing in more abstract domain models,  transitively linked to it through the \textit{parent} association.
Gluing invariants are  extremely important because they capture relationships between abstract and concrete variables during refinement that are used to demonstrate   proof obligations.

\subsection{Domain Model and Goal Model}

Each domain model is associated with a level of refinement of the \textit{SysML/KAOS} goal diagram and is likely to have as its parent, through the \textit{parent} association, another domain model (Fig. \ref{our_businessdomain_metamodel_part1}). This allows the child domain model to access and extend some elements defined in the parent domain model.	
It should be noted that the parent domain model must be associated with the refinement level of the \textit{SysML/KAOS} goal diagram directly above the refinement level to which the child domain model is associated. 
%The domain model associated with the root refinement level has no parent.

\begin{figure}[!h]
\begin{center}
\includegraphics[width=1\textwidth]{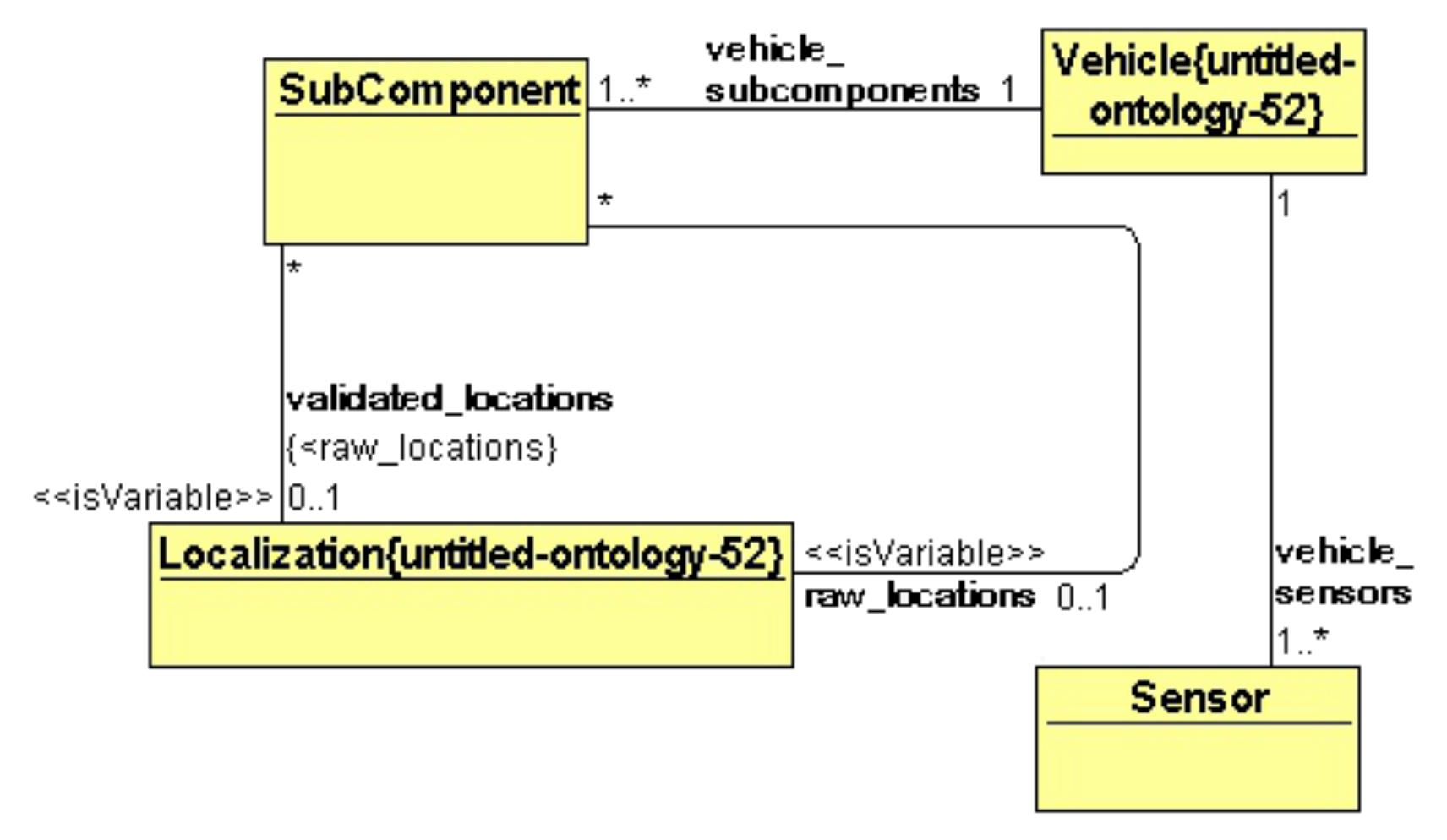}
\end{center}
\caption{\label{localization_component_shonan_book_ref1} \textit{\textbf{untitled-ontology-53}}: ontology associated to the first level of refinement}
\end{figure}

\textit{untitled-ontology-53} (Fig. \ref{localization_component_shonan_book_ref1}) has \textit{untitled-ontology-52} (Fig. \ref{localization_component_shonan_book_ref0}) as parent and defines new concepts and relationships. Each reused element is annotated with \textit{untitled-ontology-52}, the parent domain model name.
 \textit{\textbf{SubComponent}},  which is an instance of \textit{Concept}, is introduced to represent sub components of a vehicle. Each instance of \textit{Individual}  of \textit{type} \textbf{\textit{SubComponent}} associates the vehicle with a \textit{raw location}.
% Instances of \textit{Individual} of type \textbf{\textit{SubComponent}} are used to determine the \textit{raw locations} of the vehicle, modeled as an instance of \textit{Relation}.  
 \textit{\textbf{Sensor}}, which is also an instance of \textit{Concept} is introduced  to represent vehicle sensors used to validate the raw locations. Raw locations which are validated through sensors are called validated locations and are used to compute the vehicle estimated location. Each vehicle has at least one sub component and one sensor.
 \textit{untitled-ontology-54} (Fig. \ref{localization_component_shonan_book_ref2}) has \textit{untitled-ontology-53} (Fig. \ref{localization_component_shonan_book_ref1}) as parent, that allows it to manipulate  elements defined within the latter and within \textit{untitled-ontology-52} (Fig. \ref{localization_component_shonan_book_ref0}). Each reused element is annotated with the name of its domain model.
%This third abstraction level  represents the sub-components, through \textbf{\textit{SubComponent}} concept,  and the sensors, through \textbf{\textit{Sensor}} concept, of a vehicle, with the constraint that a vehicle always has at least one subcomponent and at least one sensor.
This third abstraction level  represents child concepts of \textit{SubComponent} and \textit{Sensor}.
A subcomponent is either a GPS, represented through an instance of \textit{Concept} named \textit{"Gps"}, or a Wi-Fi, represented through an instance of \textit{Concept} named \textit{"Wifi"}. A sensor is either an accelerometer, represented through an instance of \textit{Concept} named \textit{"Accelerometer"},  or a speed sensor, represented through an instance of \textit{Concept} named \textit{"SpeedSensor"}. 
Finally, \textit{\textbf{v1}} is associated to an instance of \textit{Individual} of \textit{type} \textbf{\textit{Gps}} named \textit{"g1"} and to an instance of \textit{Individual} of \textit{type}  \textbf{\textit{Wifi}} named \textit{"w1"} through \textbf{\textit{vehicle\_subcomponents}}, an instance of \textit{Relation} introduced in \textit{untitled-ontology-53}. It is also associated to  a speed sensor called \textbf{\textit{s1}} and to an accelerometer called \textbf{\textit{a1}}.

\begin{figure}[!h]
%\begin{center}
\includegraphics[width=1\textwidth]{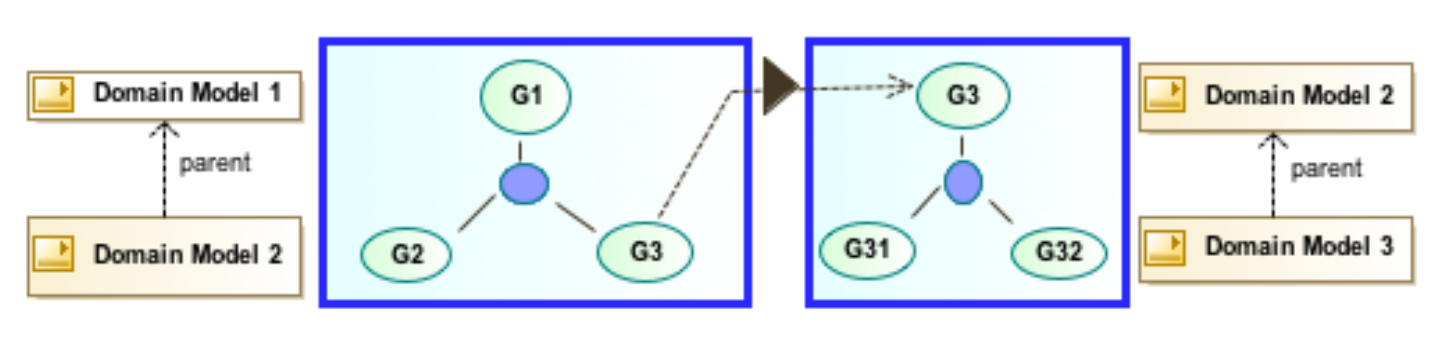}
%\end{center}
\caption{\label{SysMLKAOSMultipleDiagramHandling} Management of the partitioning of a \textit{SysML/KAOS} goal model}
\end{figure}

In order to be able to be used in the setting up of large complex systems,
\textit{SysML/KAOS} allows the refinement of a leaf   of a goal diagram in another diagram having this goal as root. For example, in Figure  \ref{SysMLKAOSMultipleDiagramHandling}, the goal \textit{G3}, which is a leaf of the first goal diagram, is the root of the second one.
When this happens, we associate to the most abstract level of the new goal diagram the domain model associated with the most concrete level of the previous goal diagram as represented in Figure  \ref{SysMLKAOSMultipleDiagramHandling}: \textit{Domain Model 2}, which is the domain model associated to the most concrete level of the first diagram, is also the domain model associated to the root of the second one. \\

\section{Conclusion}
\label{sec:6}
In this paper,  we have drawn up the  state of the art related to domain modeling in requirements engineering.  After  positioning ourselves as to the existing, we have presented our domain modeling method consisting in representing domain knowledge using an ontology modeling formalism for which a metamodel has been defined. Our approach has been illustrated through a case study dealing with a \textit{Cycab} localization  component specification.

Work in progress is aimed at developing mechanisms for the explicitness of 
\textit{SysML/KAOS} domain models semantics in \textit{Event-B}  and at 
integrating our approach within the open-source platform  \textit{Openflexo}  \cite{openflexo_link}.

\begin{acknowledgement}
%If you want to include acknowledgments of assistance and the like at the end of an individual chapter please use the \verb|acknowledgement| environment -- it will automatically render Springer's preferred layout.
This work is carried out within the framework of the  \textit{FORMOSE} project \cite{anr_FORMOSE_reference_link} funded by the French National Research Agency (ANR).
\end{acknowledgement}
%
%\section*{Appendix}
%\addcontentsline{toc}{section}{Appendix}
%
%
%//Here is appendix

\bibliographystyle{spmpsci}

\end{document}